\documentclass[lettersize,journal]{IEEEtran}
\usepackage{amsmath,amsfonts}
\usepackage{algorithm}
\usepackage[noend]{algpseudocode}
\usepackage{array}
\usepackage[caption=false,font=normalsize,labelfont=sf,textfont=sf]{subfig}
\usepackage{textcomp}
\usepackage{stfloats}
\usepackage{url}
\usepackage{xcolor}
\usepackage{verbatim}
\usepackage{graphicx}
\usepackage{dsfont}
\usepackage{hyperref}
\usepackage{cite}
\usepackage{array, makecell}
\usepackage{mathtools}
\usepackage{amsthm}
\usepackage{graphicx}
\usepackage{float}
\usepackage{amssymb}
\newtheorem{theorem}{Theorem}

\usepackage[font=small,labelfont=bf]{caption}
\newcommand\halfwidth{0.47} % use 0.47 for double column and 0.4 for single column 

\begin{document}

\title{A Speculative GLRT-Backed Approach for Robust Deep Learning-Based Array Processing}

\author{Nian-Cin Wang and Rajeev Sahay
        % <-this % stops a space

\thanks{N-C. Wang is with the Department of Data Science, University of California San Diego, San Diego, CA, 92093 USA. E-mail: \url{niw002@ucsd.edu}.}% <-this % stops a space
\thanks{R. Sahay is with the Department of Electrical and Computer Engineering, University of California San Diego, San Diego, CA, 92093 USA. E-mail: \url{r2sahay@ucsd.edu}.}% <-this % stops a space
\thanks{This work was supported in part by the University of California San Diego Academic Senate under grant RG114404 and in part by the National Science Foundation (NSF) under grant ECCS-2512912.}}



\maketitle

\begin{abstract}

Deep learning (DL) has recently emerged as an efficient approach for array processing tasks such as signal detection and direction of arrival. However, DL models lack statistical guarantees and, moreover, are highly susceptible to adversarial interference, raising security concerns about their reliability in adversarial wireless environments. In this letter, we first show that second-order statistics of the received array are spatially robust to $\ell_{p}$ bounded adversarial perturbations. Then, motivated by this theoretical result, we develop an adversarially resilient speculative array processing framework that consists of a low-latency DL classifier backed by a theoretically-grounded generalized likelihood ratio test (GLRT) validator, which operates on the spatial domain of the array, where DL is used for fast speculative inference and later confirmed with the GLRT. Empirical evaluations under multiple $\ell_{p}$ bounds, perturbation designs, and perturbation magnitudes corroborate our proposed framework and theoretical findings, demonstrating the superior performance of our proposed approach in comparison to multiple state-of-the-art baselines.

\end{abstract}


\vspace{-0.5cm}
\section{Introduction}
\IEEEPARstart{S}{ignal} detection and direction of arrival (DoA) estimation are fundamental tasks in array signal processing \cite{fundamental}. Classical approaches rely on statistical methods such as maximum likelihood estimation (MLE) and the generalized likelihood ratio test (GLRT) \cite{classical_mle}, but the required optimizations are computationally intensive, causing long delays for real-time applications.

Recent work has shown deep learning (DL) to be a low-latency alternative to MLE approaches \cite{dl_method}. However, unlike MLE, DL lacks statistical interpretability~\cite{ml_interpretation}, raising concerns on its reliability. Moreover, DL models used in wireless systems can be tricked by an attacker who transmits tiny, hard-to-notice interference signals that combine with the legitimate signal in the air, causing the DL-based receiver to make wrong decisions \cite{adv_atk_review}. Thus, neither DL nor likelihood-based approaches alone provide both interpretability and adversarial robustness for array processing.

We address this research gap by developing a robust, low-latency array processing framework that provides adversarial robustness and decision interpretability. To our knowledge, this is the first framework that jointly utilizes a DL-MLE architecture for adversarial robustness in array processing. Specifically, we consider adversarial perturbations that are bounded in magnitude in regard to (i) total (Euclidean) energy and (ii) maximum per-sample amplitude.
% Specifically, we consider $\ell_p$-bounded adversarial perturbations $\boldsymbol{\delta}$, where $\|\boldsymbol{\delta}\|_p \le \epsilon$ for a fixed budget $\epsilon$, with norm bounds of $p=2$ and $p=\infty$ as representative cases.

\textbf{Related Work:} Likelihood-based methods, such as the GLRT, have long underpinned array signal processing within a unified statistical framework \cite{classical_strength,kay1993statistical}, but incur high latency. DL has emerged as a low-latency alternative, learning directly from raw in-phase and quadrature samples or covariance measurements with strong performance across wireless tasks \cite{dl_method,chakrabarty2017cnn,shea2017dl,huang2018massive}.
However, prior work has demonstrated that DL models deployed in wireless receivers are highly vulnerable to adversarial interference. Specifically, array processing tasks have reported adversarial vulnerabilities in both DoA estimation and signal detection \cite{yang2023advdoa, li2025advdet}. Certified defenses such as adversarial training \cite{kim2022advtrain} and denoising \cite{lee2021dae} exist, but provide limited improvement and ignore statistical properties established in array processing. Alternatively, our work utilizes the GLRT’s covariance-domain, which, as we show, mitigates adversarial perturbations, offering a robust and theoretical validator for adversarial interference in array processing. Prior work has also explored using DL in conjunction with a GLRT estimator targeting nonlinear hardware distortions in signal detection only \cite{classical_weak}. In contrast, we target adversarial interference across both detection and DoA estimation, with a speculative low-latency architecture with theoretical guarantees.

\textbf{Summary of Contributions:} The main contributions of this letter are as follows:
\begin{itemize}

    \item \textbf{Spatial robustness:} We theoretically show that the second-order statistics of the received signal exhibit strong robustness to $\ell_{p}$-bounded adversarial interference.
    
    \item \textbf{Speculative array processing framework:} We develop and empirically demonstrate a speculative framework that integrates a DL classifier with a GLRT estimator, which operates on the spatial domain, to improve robustness against adversarial interference in array processing.
    
    % \item \textbf{Empirical evaluation:} We empirically demonstrate the efficacy of our speculative framework on two array processing tasks: signal detection and direction of arrival. % over multiple adversarial interference designs and perturbation magnitudes and demonstrate its superiority over multiple baselines. 
\end{itemize}

\vspace{-0.3cm}
\section{Methodology}

\subsection{Signal Modeling}
\label{sec:signalmodel}

We consider a uniform linear array consisting of $M$ sensors, equally spaced by $d$ wavelengths. 
At snapshot $t$, the received antenna array observation is given by
\begin{equation}
    \mathbf{z}(t) = \mathbf{a}(\theta_1)s_1(t) + \mathbf{a}(\theta_2)s_2(t) + \mathbf{n}(t), \quad t = 1,\dots,T,
\end{equation}
where $\mathbf{z}(t) \in \mathbb{C}^M$ is the $M$-dimensional array output, 
$\mathbf{a}(\theta_{1})$ and $\mathbf{a}(\theta_{2})$ are the steering vectors corresponding to the narrow-band DoA $\theta_{1}$ and $\theta_{2}$, respectively,  $s_1(t)$ and $s_2(t)$ denote the complex amplitudes of the interference signal and the signal of interest (SOI), respectively, 
$\mathbf{n}(t) \sim \mathcal{CN}(\mathbf{0},\sigma^2 \mathbf{I})$ is additive white Gaussian noise, and $T$ is the number of temporal snapshots. 
% Both $\theta_1$ and $\theta_2$ are restricted to narrow-band angles, which also defines the range of DoAs considered in estimation. 
The received data across $T$ snapshots is collected in the matrix $\mathbf{Z} = [\mathbf{z}(1), \mathbf{z}(2), \dots, \mathbf{z}(T)] \in \mathbb{C}^{M \times T}$. Given $\mathbf{Z}$, we focus on two array processing tasks:
\begin{itemize}
    \item \textbf{Signal detection:} Here, the objective is to determine whether the SOI is present in $\mathbf{Z}$. 
    This is formulated as a binary hypothesis test. 
    We define $H_0$ as the null hypothesis where only the interference $s_1(t)$ and noise are present (i.e., $s_2(t)=0 \hspace{1mm} \forall \hspace{1mm} t$), 
    and $H_1$ as the alternative hypothesis where both $s_1(t)$ and the SOI $s_2(t)$ are present (i.e., $s_2(t) \neq 0 \hspace{1mm} \forall \hspace{1mm} t$). Formally, the distribution of $\mathbf{Z}$ under each hypothesis is given by
    % \begin{equation}
    %     \begin{cases}
    %         H_0: \mathbf{z}(t) = \mathbf{a}(\theta_1)s_1(t) + \mathbf{n}(t), \\
    %         H_1: \mathbf{z}(t) = \mathbf{a}(\theta_1)s_1(t) + \mathbf{a}(\theta_2)s_2(t) + \mathbf{n}(t).
    %     \end{cases}
    % \end{equation}
    \begin{equation}
        \begin{cases}
            H_0: \mathbf{Z} \sim  \mathcal{CN}(\mathbf{0},\mathbf{R}) \\
            H_1: \mathbf{Z} \sim \mathcal{CN}(\mathbf{AST},\mathbf{R}), 
        \end{cases}
    \end{equation}
    where $\mathbf{R}$ is the true unknown covariance matrix of $\mathbf{Z}$, $\mathbf{A} = [\mathbf{a}(\theta_{2})] \in \mathbb{C}^{M \times 1}$ is the antenna array response of the SOI (since we are exclusively considering a receiver with one SOI), $\mathbf{S} \in \mathbb{C}^{1 \times S}$ is the row vector of signal amplitudes, $\mathbf{T} \in \mathbb{C}^{S \times L}$ is the matrix $[\mathbf{0} \quad \mathbf{I}_{S}]$, and $S = L - t_{0}$, where $t_{0}$ denotes the time sample at which $s_{2}$ becomes active in $\mathbf{Z}$.

    % Here, we consider the case where there is at most one SOI. 
    % The received data matrix follows $\mathbf{Z} \sim \mathcal{CN}(\mathbf{0},\mathbf{R})$ under $H_0$ 
    % and $\mathbf{Z} \sim \mathcal{CN}(\mathbf{A}s_2,\mathbf{R})$ under $H_1$, 
    % where $\mathbf{R}$ is the unknown covariance matrix and $\mathbf{A} = \mathbf{a}(\theta_2)$ denotes the steering vector of the SOI. 
    % We consider the scenario where, when the SOI is present, it is introduced at a random activation time $t_0$, 
    % with $T$ being the number of snapshots, and assigned a randomly selected DoA $\theta_2 \in \Theta$, where $\Theta$ is the set of candidate DoA angles.

    \item \textbf{DoA estimation:} In this case, the SOI is always present (i.e., $H_1$ holds), 
    and the task reduces to estimating its DoA $\theta_2 \in \Theta$, where $\Theta$ is the set of candidate DoA angles. ere, $H_1$ is assumed true (from signal detection), with DoA estimation performed as a subsequent step. %This transforms the task from a binary detection into a multi-class classification problem, where we are interested in estimating $\theta_2$ given $\mathbf{Z}$ and that $H_{1}$ holds.
\end{itemize}

\vspace{-0.25cm}
\subsection{GLRT Formulation}
\label{sec:glrt}
The GLRT provides a statistical framework for both signal detection and DoA estimation \cite{kay1993statistical,kelly1986glrt}. 
Let $h_0(\mathbf{Z};\mathbf{R})$ and $h_1(\mathbf{Z};\mathbf{R},\mathbf{S})$ denote the complex Gaussian likelihood density functions of the received data $\mathbf{Z}$ under hypotheses $H_0$ and $H_1$, respectively. Then, the GLRT statistic, by definition, is given by
\begin{equation}
    \Lambda(\mathbf{Z}) \triangleq 
    \frac{\max_{\mathbf{R},\,\mathbf{S}} \; h_1(\mathbf{Z}; \mathbf{R}, \mathbf{S})}
         {\max_{\mathbf{R}} \; h_0(\mathbf{Z}; \mathbf{R})},
\end{equation}
and we declare $H_1$ whenever $\Lambda(\mathbf{Z}) \geq \gamma$. 
Maximization with respect to $\mathbf{R}$ yields the empirical covariance, given by
\begin{equation}
    \hat{\mathbf{R}}_0 = \mathbf{Z}\mathbf{Z}^{H}, \qquad
    \hat{\mathbf{R}}_1 = (\mathbf{Z} - \mathbf{AST})(\mathbf{Z} - \mathbf{AST})^{H},
\end{equation}
where $\hat{\mathbf{R}}_0$ and $\hat{\mathbf{R}}_1$ are the maximum-likelihood covariance estimates under $H_0$ and $H_1$. Substituting these back into the likelihoods, the GLRT reduces to a ratio of determinants and can be equivalently expressed as
\begin{equation}
    \Lambda(\mathbf{Z}) = 
    \frac{|\mathbf{Z}\mathbf{Z}^{H}|}
         {\min_{\mathbf{S}} |(\mathbf{Z} - \mathbf{AST})(\mathbf{Z} - \mathbf{AST)}^{H}|},
\end{equation}
where $(\cdot)^H$ denotes the Hermitian transpose and $|\cdot|$ denotes the determinant. 
From here, we can minimize over $\mathbf{S}$ (see \cite{kelly1986glrt} for full derivation) to arrive at 
\begin{equation} \label{glrt2} 
    \Lambda(\mathbf{Z}) = \frac {|\mathbf{A}^{\text{H}} (\mathbf{Z}_{\text{old}} \mathbf{Z}_{\text{old}}^{\text{H}})^{-1} \mathbf{A}|} {|\mathbf{A}^{\text{H}} (\mathbf{Z} \mathbf{Z}^{\text{H}})^{-1} \mathbf{A}|},
\end{equation}
where we have introduced the empirical quantity $\mathbf{Z}_{\text{old}}$, which is a partition of subsequent time samples in $\mathbf{Z}$. We will define $\mathbf{Z}_{\text{old}}$ for each array processing task below: % in Secs. \ref{sec:glrt_det} and \ref{sec:glrt_doa}. % This general framework can be specialized depending on the task: 
% In this work, we consider two cases: 
% (i) GLRT for detection, where the SOI turn-on time is unknown, and 
% (ii) GLRT for DoA estimation, where the SOI is known to activate at a fixed time and we aim to classify $\theta_{2} \in \Theta$.

\subsubsection{GLRT for Detection} \label{sec:glrt_det}
For detection, we apply the GLRT over consecutive blocks of $k$ time samples. 
Specifically, we define two length-$k$ partitions of $\mathbf{Z}$, denoted by % $\mathbf{Z}_{\text{old}} = [\,\mathbf{z}(i), \mathbf{z}(i+1), \cdots, \mathbf{z}(i+k)\,]$ and $\mathbf{Z}_{\text{new}} = [\,\mathbf{z}(i+k+1), \mathbf{z}(i+k+2), \ldots, \mathbf{z}(i+2k)\,]$
\begin{equation}
\label{eq:det_window}
\begin{aligned}
\mathbf{Z}_{\text{old}} &= [\,\mathbf{z}(i), \mathbf{z}(i+1), \cdots, \mathbf{z}(i+k)\,], \\
\mathbf{Z}_{\text{new}} &= [\,\mathbf{z}(i+k+1), \mathbf{z}(i+k+2), \ldots, \mathbf{z}(i+2k)\,]
\end{aligned}
\end{equation}
for $i = 1, \dots, T-k$. 
Each partition $\mathbf{Z}_{\text{old}}$ and $\mathbf{Z}_{\text{new}}$ therefore consists of two non-overlapping windows shifted forward by one snapshot. 
By iterating over all $i$, the GLRT is reevaluated across the entire sequence of $T$ snapshots, ensuring that potential activations of the SOI at arbitrary time indices are captured. In practice, we compute the corresponding empirical covariance matrices as
\begin{equation}
\label{eq:covariance_estimate}
    \hat{\mathbf{R}}_{\text{old}} = \mathbf{Z}_{\text{old}}\mathbf{Z}_{\text{old}}^{H}, \qquad
    \hat{\mathbf{R}}_{\text{new}} = \mathbf{Z}_{\text{new}}\mathbf{Z}_{\text{new}}^{H},
\end{equation}
apply diagonal loading before inversion, i.e., invert $\hat{\mathbf{R}}_{\text{old}} + \beta \mathbf{I}$, with $\beta>0$ small, and apply the monotonically related (to (\ref{glrt2})) test statistic
\begin{equation} \label{det_stat}
    T_i = \mathrm{tr}\!\left((\hat{\mathbf{R}}_{\text{old}} + \beta \mathbf{I})^{-1} \hat{\mathbf{R}}_{\text{new}}\right),
\end{equation}
where $\text{tr}(\cdot)$ is the trace of $\cdot$, with $\beta \mathbf{I}$ applied analogously to $\hat{\mathbf{R}}_{\text{old}}^{-1}$ in \eqref{glrt_doa} for DoA. We adopt $T_i$ as our detection statistic, declaring $H_1$ whenever $T_i \geq \gamma_T$. The threshold $\gamma_T$ is calibrated empirically on held-out $H_0$-only data, set to the value of $T_i$ achieving a target false-alarm probability of 0.05, consistent with standard Neyman-Pearson threshold selection.

We denote the resulting GLRT detection decision as
\begin{equation}
\label{glrt_det}
    \hat{y}^{\mathrm{GLRT}}_{\mathrm{det}} =
    \begin{cases}
        1, & T_i \ge \gamma_T, \\
        0, & T_i < \gamma_T.
    \end{cases}
\end{equation}

\subsubsection{GLRT for DoA Estimation} \label{sec:glrt_doa}
For DoA estimation, we assume $H_1$ holds and the SOI activates at $t = t_0$. 
We form two adjacent windows of length $k$ around $t_0$ given by %$\mathbf{Z}_{\text{old}} = [\mathbf{z}(t_0-k), \dots, \mathbf{z}(t_0-1)]$ and $\mathbf{Z}_{\text{new}} = [\mathbf{z}(t_0), \dots, \mathbf{z}(t_0+k-1)]$
\begin{equation}
\label{eq:doa_window}
\begin{aligned}
\mathbf{Z}_{\text{old}} = [\mathbf{z}(t_0-k), \dots, \mathbf{z}(t_0-1)], \\
\mathbf{Z}_{\text{new}} = [\mathbf{z}(t_0), \dots, \mathbf{z}(t_0+k-1)],
\end{aligned}
\end{equation}
and compute their empirical covariances using the expression in ~\eqref{eq:covariance_estimate}.
We then use a matched filter to identify $\theta_{2}$ by forming $\mathbf{M} = (\hat{\mathbf{R}}_{\text{old}} + \beta \mathbf{I})^{-1}\hat{\mathbf{R}}_{\text{new}}$, 
extracting its maximum eigenvector $\mathbf{v}_{\max}$, 
and estimating the DoA over the steering grid $\Theta$. We denote this GLRT DoA estimate by 
\begin{equation} 
\label{glrt_doa}
    \hat{y}^{\mathrm{GLRT}}_{\mathrm{doa}} = 
    \arg\max_{\theta \in \Theta} 
    \big|\mathbf{a}(\theta)^{H} \mathbf{v}_{\max}\big|.
\end{equation}

\vspace{-0.65cm}
\subsection{Data-Driven Classifier Modeling}
\label{sec:cnn}
Although (\ref{glrt_det}) and (\ref{glrt_doa}) provide theoretically grounded test statistics, they are computationally costly: (\ref{glrt_det}) requires partitions of $\mathbf{Z}$ that grow with the observation window $T$, and (\ref{glrt_doa}) requires computing the maximum eigenvector over all candidate angles via a matched filter. To address this, we now consider a DL classifier trained to infer the underlying hypothesis or DoA directly from the received array data.

For the signal detection task, we define the detection classifier as $f_{\mathrm{det}}(\cdot\,;\phi_{\mathrm{det}}): \mathbb{R}^{M\times T\times 2} \rightarrow \{0,1\}$ parameterized by $\phi_{\mathrm{det}}$, where the real and imaginary components of $\mathbf{Z} \in \mathbb{R}^{M \times T \times 2}$ construct the input. The classifier, thus, produces the speculative detection output
\begin{equation}
\label{eq:cnn_det}
    \hat{y}^{\mathrm{DL}}_{\mathrm{det}} 
    = f_{\mathrm{det}}(\mathbf{Z};\phi_{\mathrm{det}}),
    \quad 
    \hat{y}^{\mathrm{DL}}_{\mathrm{det}} \in \{0,1\}.
\end{equation}

From (\ref{glrt_doa}), we see that DoA estimation is achieved on second order statistics. Thus, our DL DoA classifier operates on the empirical covariance of the received signal. Specifically, we define the DoA classifier as $f_{\mathrm{doa}}(\cdot;\phi_{\mathrm{doa}}):
\mathbb{R}^{M \times 2M \times 2} \rightarrow \Theta$ parameterized by $\phi_{doa}$, where the real and imaginary components of $[\hat{\mathbf{R}}_{\text{old}}, \hat{\mathbf{R}}_{\text{new}}] \in \mathbb{R}^{M \times 2M \times 2}$ construct the input. The DoA classifier, thus produces the speculative angle prediction 
\begin{equation}
\label{eq:cnn_doa}
    \hat{y}^{\mathrm{DL}}_{\mathrm{doa}} 
    = f_{\mathrm{doa}}([\hat{\mathbf{R}}_{\text{old}}, \hat{\mathbf{R}}_{\text{new}}]; \phi_{\mathrm{doa}}),
    \quad
    \hat{y}^{\mathrm{DL}}_{\mathrm{doa}} \in \Theta.
\end{equation}

\vspace{-0.6cm}
\subsection{Adversarial Attacks}
\label{sec:adv_attacks}

Despite their efficiency, DL classifiers are susceptible to adversarial interference, which is specifically crafted to induce misclassification on DL models at the receiver. To craft an adversarial attack, denoted by $\boldsymbol{\delta}$, an adversary designs a perturbation to maximize the classifier’s loss. The received adversarial signal is given by $\tilde{\mathbf{Z}} = \mathbf{Z} + \boldsymbol{\delta}$. In this work, we consider two adversarial interference methods: the fast gradient sign method (FGSM) \cite{goodfellow2015fgsm} and projected gradient descent (PGD) \cite{madry2019adv}. We consider each of these attacks under two $\ell_p$ norms of $\boldsymbol{\delta}$: $p = 2$ and $p = \infty$. Here, $\boldsymbol{\delta}$ is the additive perturbation applied to $\mathbf{Z}$ (or to $[\hat{\mathbf{R}}_{\text{old}}, \hat{\mathbf{R}}_{\text{new}}]$ for DoA), and $p$ indexes its norm constraint, with $p=2$ Euclidean-bounded and $p=\infty$ bounded in maximum per-entry magnitude.

\subsubsection{FGSM}
FGSM is a single-step gradient attack that exhausts the power budget in one iteration. Here, $\boldsymbol{\delta} = \epsilon \,
    \mathrm{sign}\!\big(\nabla_{\mathbf{Z}} \mathcal{L}(\mathbf{Z}, y; \phi)\big)$
when $p = \infty$, where $\epsilon = ||\boldsymbol{\delta}||_{\infty}$ controls the perturbation magnitude, $y$ is the true label, and $\phi$ is either $\phi_{\det}$ or $\phi_{\text{doa}}$ depending on whether the perturbation is crafted for detection or DoA, respectively. For $\ell_2$-bounded attacks, the gradient can be normalized and calculated according to $\boldsymbol{\delta} = \epsilon \nabla_{\mathbf{Z}} \mathcal{L}(\mathbf{Z}, y; \phi) / \|\nabla_{\mathbf{Z}} \mathcal{L}(\mathbf{Z}, y; \phi)\|_2$.

\subsubsection{PGD}
PGD is an iterative extension of FGSM. Instead of applying the entire $\varepsilon$-bounded change in a single update, PGD applies smaller increments of size $\alpha = \varepsilon / Q$ over $Q$ iterations, recalculating the gradient after each 
update. An $\ell_{p}$-bounded PGD perturbation on iteration $q$ is given by
\begin{equation}
    \mathbf{Z}^{(q+1)} = \Pi_{B_p(\mathbf{Z}^{(0)}, \epsilon)}\big(\mathbf{Z}^{(q)} + \alpha\cdot \boldsymbol{\Delta}_{p}(\mathbf{Z}^{(q)}) \big), 
\end{equation}
where $\Pi_{B_p(\mathbf{Z}^{(0)}, \epsilon)}(\cdot)$ is the projection of $\cdot$ onto the $\ell_{p}$ norm ball centered at $\mathbf{Z}^{(0)}$ with radius $\epsilon$, $\mathbf{Z}^{(0)} = \mathbf{Z}$, 
\begin{equation}
    \boldsymbol{\Delta}_{\infty}(\mathbf{Z}^{(q)}) = \text{sign} \big(\nabla_{\mathbf{Z}^{(q)}} \mathcal{L}(\mathbf{Z}^{(q)}, y; \phi)\big),
\end{equation}
when $p = \infty$, and
\begin{equation} \label{last}
    \boldsymbol{\Delta}_{2}(\mathbf{Z}^{(q)}) = \frac{\nabla_{\mathbf{Z}^{(q)}} \mathcal{L}(\mathbf{Z}^{(q)}, y; \phi)}
         {\|\nabla_{\mathbf{Z}^{(q)}} \mathcal{L}(\mathbf{Z}^{(q)}, y; \phi)\|_2},
\end{equation}
when $p = 2$. The final perturbation is given by $\tilde{\mathbf{Z}} = \mathbf{Z}^{(Q)}$.

\vspace{-0.4cm}
\subsection{Speculative Processing}
\label{sec:proposed}

We now analytically characterize the impact of adversarial interference on the second order statistics of $\mathbf{Z}$.

\begin{theorem}
\label{thm:cov-stability}
For any adversarial perturbation $\boldsymbol{\delta}$ satisfying $\|\boldsymbol{\delta}\|_p \le \epsilon$, where $\epsilon$ $\ll \|\mathbf{Z}\|_2$, the energy induced by the additive interference is bounded according to
\begin{equation}
    ||S(\mathbf{Z} + \boldsymbol{\delta}) - S(\mathbf{Z})||_{2} \leq C_{1}\epsilon + C_{2}\epsilon^2
    \qquad \forall\, p \in [1,\infty],
\end{equation}
where $S(\cdot)$ denotes the covariance of $\cdot$, and $C_1$ and $C_2$ depend on $p$, $T$, and $\|\mathbf{Z}\|_2$. In particular, the relative deviation becomes small when the perturbation energy is small compared to the signal energy. % Thus, we see that the deviation induced by $\boldsymbol{\delta}$ is negligible in the spatial domain, reducing the impact of $\ell_{p}$-bounded adversarial interference. 
\end{theorem}
\vspace{-0.3cm}
\begin{proof} See Appendix A. \end{proof}
\vspace{-0.2cm}
\textbf{Remark:} Theorem~\ref{thm:cov-stability} bounds the perturbation of $S(\mathbf{Z})$, whereas the GLRT statistic is computed from $\hat{\mathbf{R}}_{\text{old}}^{-1}\hat{\mathbf{R}}_{\text{new}}$ (or from $\mathbf{v}_{\max}$ for DoA). For $D = S(\mathbf{Z}+\boldsymbol{\delta}) - S(\mathbf{Z})$, standard perturbation results give an inverse perturbation of $\mathcal{O}(\|\hat{\mathbf{R}}_{\text{old}}^{-1}\|_2^2\|D\|_2)$, and, via Davis--Kahan, an eigenvector perturbation scales with $\|D\|_2$ over the eigenvalue gap, where both terms grow with the condition number of $\hat{\mathbf{R}}_{\text{old}}$. Applying diagonal loading, as discussed in Sec. \ref{sec:glrt_det}, keeps $\hat{\mathbf{R}}_{\text{old}}$ well-conditioned and ensures Theorem~\ref{thm:cov-stability}'s small covariance perturbation yields a correspondingly small perturbation of the GLRT statistic, which relies on $\hat{\mathbf{R}}_{\text{old}}^{-1}$ and $\mathbf{v}_{\max}$. Thus, we see that the robustness of the GLRT rises from its use of the covariance through a well-conditioned matrix inverse, whose sensitivity to such small perturbations is explicitly bounded. A DL classifier trained on the same covariance input is not guaranteed this robustness, since its nonlinear decision boundaries amplify small covariance perturbations, making it susceptible to adversarial interference.

\begin{algorithm}[t] % use t for double column and H for single column
\caption{Speculative Inference Framework}
\label{alg:cnn-glrt}
\begin{algorithmic}[1]
\Require $\mathbf{Z}$, 
task (det or doa), 
$f_{\mathrm{det}}(\cdot; \phi_{\text{det}})$, $f_{\mathrm{doa}}(\cdot; \phi_{\text{doa}})$, 
$\Theta$
% \Ensure Final output $\hat{y}_{\mathrm{det}}$ or $\hat{y}_{\mathrm{doa}}$

\State \textbf{Speculative Stage (DL classifier):}
\If{task = det}
    \State $\hat{y}^{\mathrm{DL}}_{\mathrm{det}} \gets f_{\mathrm{det}}(\mathbf{Z}_{\text{received}})$
\Else 
    \State $\hat{y}^{\mathrm{DL}}_{\mathrm{doa}} \gets f_{\mathrm{doa}}(\mathbf{Z}_{\text{received}})$
\EndIf

\State Continue post processing (e.g., decoding, etc.) using the speculative DL classifier output
\Statex{\hspace{0.5em}\textbf{--- GLRT validation runs asynchronously ---}}

\If{task = det}
    \State Compute $\hat{\mathbf{R}}_{\mathrm{old}}$ and $\hat{\mathbf{R}}_{\mathrm{new}}$ using \eqref{eq:det_window} and \eqref{eq:covariance_estimate} 
    \State $T_i \gets \mathrm{tr}((\hat{\mathbf{R}}_{\mathrm{old}} + \beta\mathbf{I})^{-1}\hat{\mathbf{R}}_{\mathrm{new}})$
    % \State $\hat{y}^{\mathrm{GLRT}}_{\mathrm{det}} \gets \mathbb{1}[\max_i T_i \ge \gamma_T]$
    \State $\hat{y}^{\mathrm{GLRT}}_{\mathrm{det}} \gets 1$ if $\max_i T_i \ge \gamma_T$, else $0$
    \State \textbf{Consistency Check:} set $\hat{y}_{\mathrm{det}} \gets \hat{y}^{\mathrm{DL}}_{\mathrm{det}}$ if $\hat{y}^{\mathrm{DL}}_{\mathrm{det}} = \hat{y}^{\mathrm{GLRT}}_{\mathrm{det}}$, else $\hat{y}^{\mathrm{GLRT}}_{\mathrm{det}}$

\Else \Comment{task = doa}
    \State Compute $\hat{\mathbf{R}}_{\mathrm{old}}$ and $\hat{\mathbf{R}}_{\mathrm{new}}$ using \eqref{eq:covariance_estimate} and \eqref{eq:doa_window}
    \State $\mathbf{M} \gets (\hat{\mathbf{R}}_{\mathrm{old}} + \beta\mathbf{I})^{-1}\hat{\mathbf{R}}_{\mathrm{new}}$
    \State $\mathbf{v}_{\max} \gets$ dominant eigenvector of $\mathbf{M}$
    \State $\hat{y}^{\mathrm{GLRT}}_{\mathrm{doa}} \gets \arg\max_{\theta \in \Theta} |\mathbf{a}(\theta)^H \mathbf{v}_{\max}|$
    \State \textbf{Consistency Check:} set $\hat{y}_{\mathrm{doa}} \gets \hat{y}^{\mathrm{DL}}_{\mathrm{doa}}$ if $\hat{y}^{\mathrm{DL}}_{\mathrm{doa}} = \hat{y}^{\mathrm{GLRT}}_{\mathrm{doa}}$, else $\hat{y}^{\mathrm{GLRT}}_{\mathrm{doa}}$
\EndIf

\State \Return $\hat{y}_{\mathrm{det}}$ or $\hat{y}_{\mathrm{doa}}$
\end{algorithmic}
\end{algorithm}

Motivated by Theorem \ref{thm:cov-stability}, we propose a speculative array processing framework: the DL classifier provides low-latency inference on a (potentially adversarial) array, and post-processing (demodulation, decoding, etc.) proceeds immediately using this prediction. Simultaneously, the GLRT validates the same signal, taking longer due to its higher computational cost. However, the receiver is not stalled while waiting, since post-processing already proceeds with the DL result. Once the GLRT completes, a consistency check confirms the DL decision if the two agree, or restarts postprocessing with the GLRT output if they disagree. This combines the speed of DL inference with the statistical reliability of the GLRT through asynchronous validation. Our complete framework is shown in Algorithm \ref{alg:cnn-glrt}.

Theorem \ref{thm:cov-stability} is a worst-case bound holding for any perturbation (gradient-based or non-gradient-based) $\boldsymbol{\delta}$ satisfying $\|\boldsymbol{\delta}\|_p \le \epsilon$, consistent with the over-the-air adversarial interference threat models considered in the wireless literature \cite{adv_atk_review,yang2023advdoa}. Consequently, our framework remains robust to any adaptive adversary that directly targets the GLRT statistic or covariance estimate via such a bounded perturbation, and since Algorithm \ref{alg:cnn-glrt}'s consistency check always defers to the GLRT output on disagreement, an adversary targeting this mechanism cannot corrupt the correctness of the final decision. However, an adversary that shifts the distribution of $\mathbf{Z}$ away from that assumed by the GLRT, or that directly manipulates the covariance estimates rather than injecting $\boldsymbol{\delta}$, would break the GLRT itself; such scenarios are outside the scope of this letter. Instead, given a statistically valid GLRT, our letter shows that any adversarial attack with $\|\boldsymbol{\delta}\|_p \le \epsilon$ is bounded (Theorem 1), and our approach is robust to such attacks.

Under an attack that always causes disagreement, Algorithm~\ref{alg:cnn-glrt}'s worst-case latency is $\tau_{\mathrm{DL}} + \tau_{\mathrm{GLRT}} > \tau_{\mathrm{GLRT}}$, where $\tau_{\mathrm{DL}} \ll \tau_{\mathrm{GLRT}}$. Asymptotically, $\tau_{\mathrm{GLRT}} = \mathcal{O}((T-k)M^3)$ (repeated $M\times M$ covariance LU inversion across sliding windows) dominates $\tau_{\mathrm{DL}} = \mathcal{O}(MT)$ (a single CNN forward pass), so $\tau_{\mathrm{DL}} + \tau_{\mathrm{GLRT}} = \mathcal{O}(\tau_{\mathrm{GLRT}})$. Thus, we have the same asymptotic order for Algorithm \ref{alg:cnn-glrt} as the standalone GLRT.

\vspace{-0.5cm}
\section{Performance Evaluation}
\label{sec:results}
\vspace{-0.35cm}
\subsection{Experimental Setup}
\label{sec:exp_setup}
\vspace{-0.15cm}

We generate two datasets using MATLAB's \textit{Phased Array System Toolbox}, with $M = 8$, $d = \lambda/2$, $\beta=1e-6$, and unit-energy signals. For detection, 30,000 signals with $T = 500$ are evenly split between (0) $s_1$ only and (1) $s_1$ and $s_2$ (the SOI), with $\theta_1,\theta_2$ uniform on $[-60^{\circ}, 60^{\circ}]$ and $s_2$ activated at a random time $t_0$. For DoA, 30,000 samples with $T = 1500$ each contain $s_1$ and $s_2$ (with $s_{2}$ activated at $t_0=T/2$), with $\theta_2$ uniform on $[-60^{\circ}, 60^{\circ}]$ in $2^{\circ}$ steps (61 classes).

All signals use a fixed signal-to-noise ratio (SNR) of $\mathrm{SNR} = 0$ dB, held constant across varying PSR levels, so accuracy degradation reflects adversarial interference rather than the noise floor. We quantify the adversarial potency using the perturbation-to-signal ratio (PSR), which provides a norm-independent measure of distortion. The PSR in relation to $\boldsymbol{\delta}$ is defined as $\mathrm{PSR\,[dB]} =
10 \log_{10} \left(\mathbb{E}[\|\boldsymbol{\delta}\|_2^2] / \mathbb{E}[\|\mathbf{Z}\|_2^2] \right) \text{[dB]}$.

The efficacy of our framework is measured in terms of classification accuracy: the fraction of test signals whose predicted label ($\hat{y}_{\mathrm{det}} \in \{0,1\}$ for detection, or $\hat{y}_{\mathrm{doa}} \in \Theta$ 61 classes for DoA) matches ground truth, evaluated at each PSR level. The GLRT was implemented using $k=10$ and $k=750$ for signal detection and DoA, respectively. For DL, we implement a convolutional neural network (CNN) for each task, both trained with Adam, categorical cross-entropy loss, learning rate 0.001, and batch size 32 (determined via grid search). The architecture of each CNN is given in Table~\ref{tab:cnn_architecture_binary}.

\begin{table}[t]
\centering
\caption{CNN architectures for each array processing task.}
\label{tab:cnn_architecture_binary}
\begin{tabular}{| c | c | c | c |}
\hline
\textbf{Layer} & \textbf{Activation} & \textbf{Shape (Det.)} & \textbf{Shape (DoA)} \\
\hline
Conv 1 & ReLU & $3\times3\times32$ & $2\times2\times32$ \\
BatchNorm 1 & - & -- & $2\times2\times32$ \\
MaxPool 1 & - & $2\times2\times32$ & $1\times2\times32$ \\
Conv 2 & ReLU & $3\times3\times64$ & $2\times2\times64$ \\
BatchNorm 2 & - & -- & $2\times2\times64$ \\
MaxPool 2 & - & $1\times2\times64$ & $1\times2\times64$ \\
Flatten & - & -- & -- \\
Dense 1 (Dropout) & ReLU & 128 (50\%) & 128 (30\%) \\
Output & Softmax & 2 & 61 \\
\hline
\end{tabular}
\end{table}

\vspace{-0.5cm}
\subsection{Simulation Results}

\begin{figure}[ht!]
    \centering
    % First row of subfigures
    \subfloat[FGSM $\ell_\infty$]{%
        \includegraphics[width=\halfwidth\columnwidth]{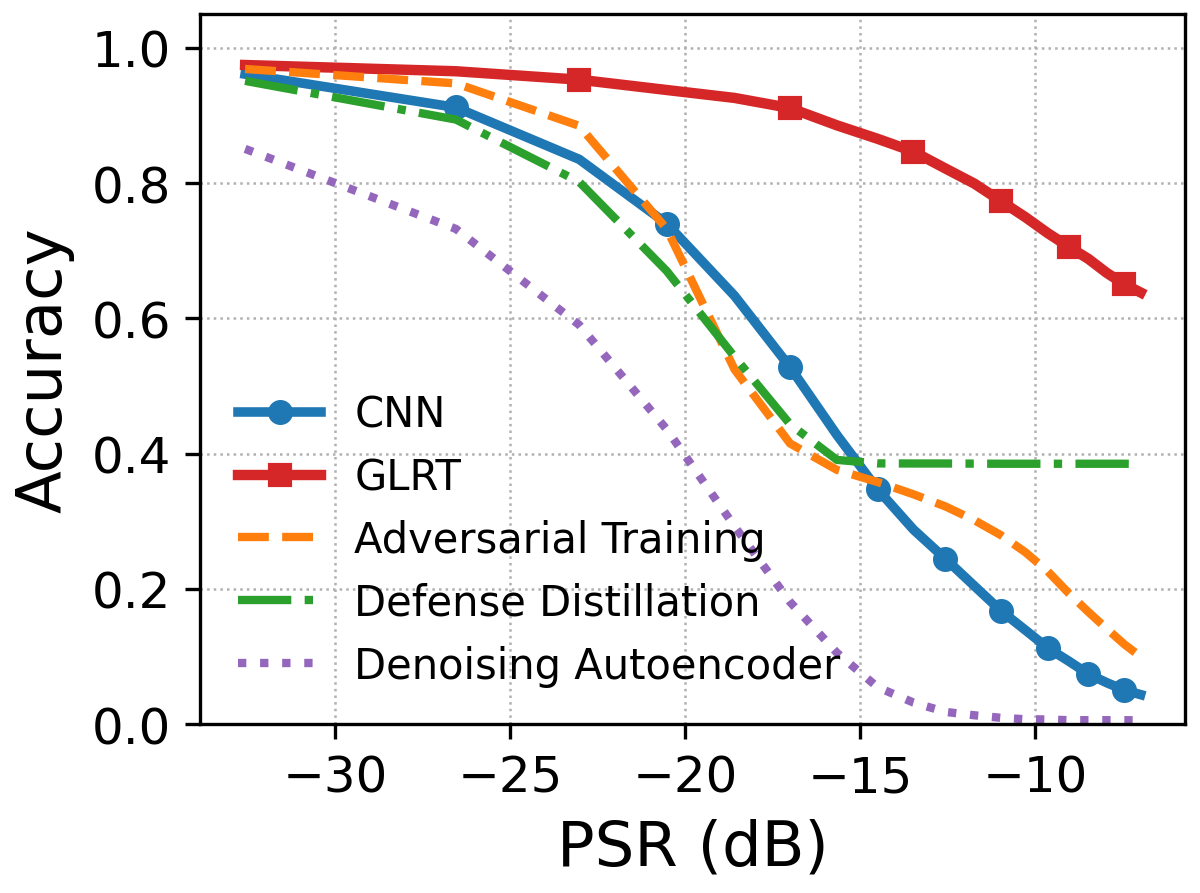}
    }
    \subfloat[FGSM $\ell_2$]{%
        \includegraphics[width=\halfwidth\columnwidth]{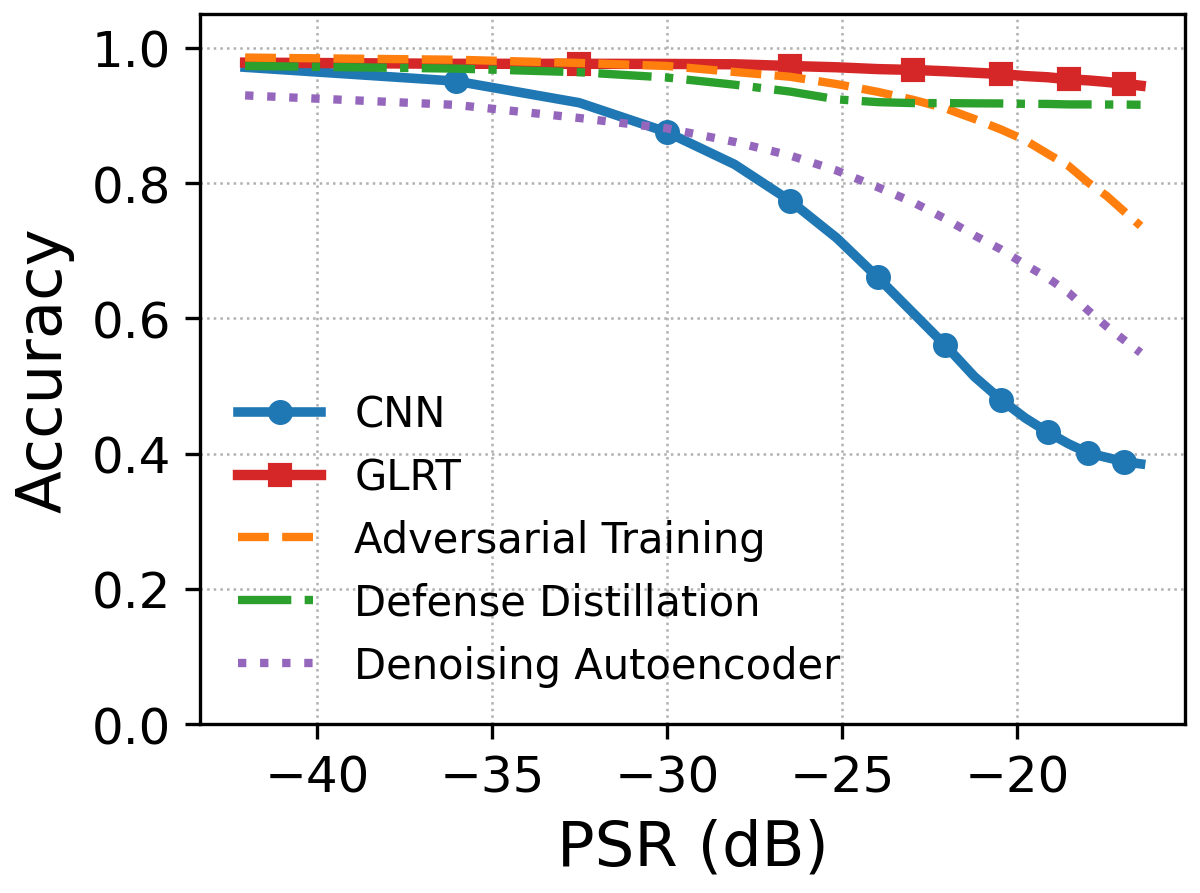}
    }
    
    % Second row of subfigures
    \subfloat[PGD $\ell_\infty$]{%
        \includegraphics[width=\halfwidth\columnwidth]{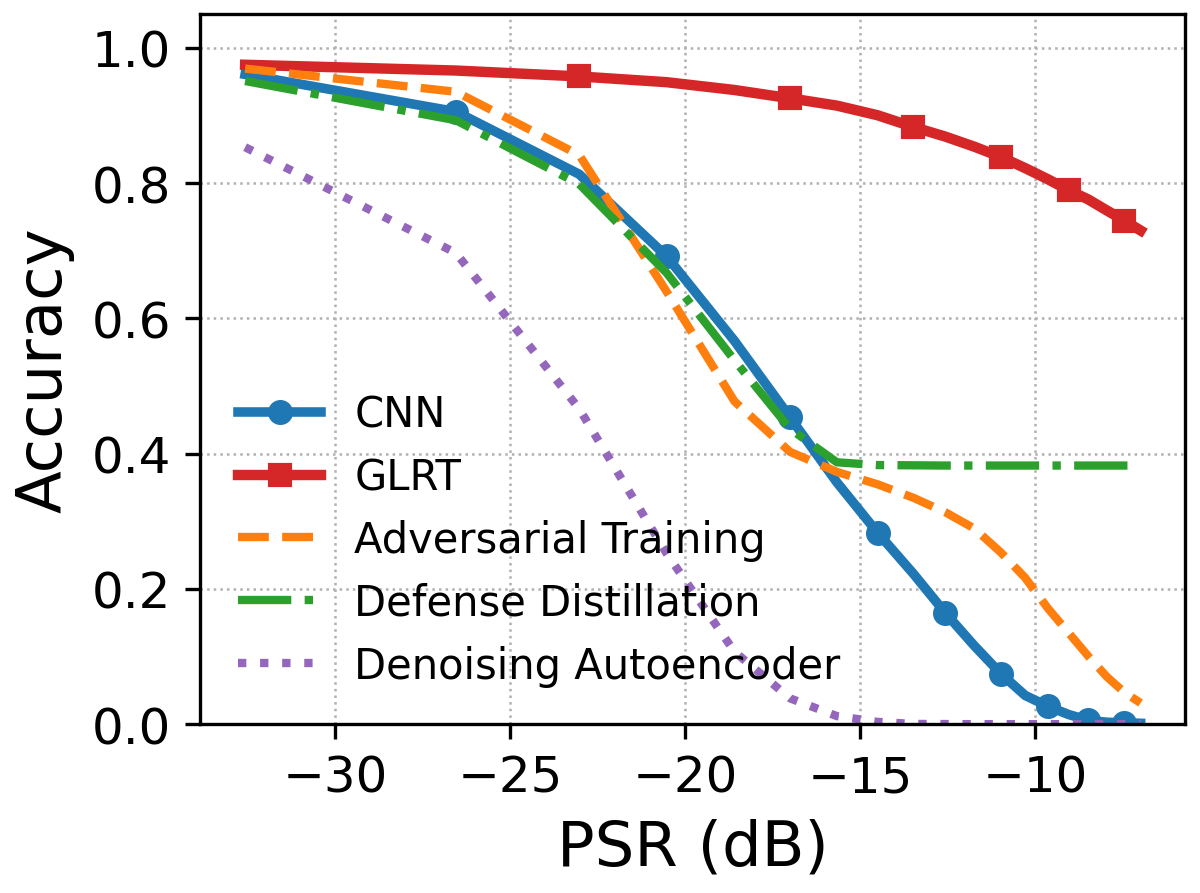}
    }
    \subfloat[PGD $\ell_2$]{%
        \includegraphics[width=\halfwidth\columnwidth]{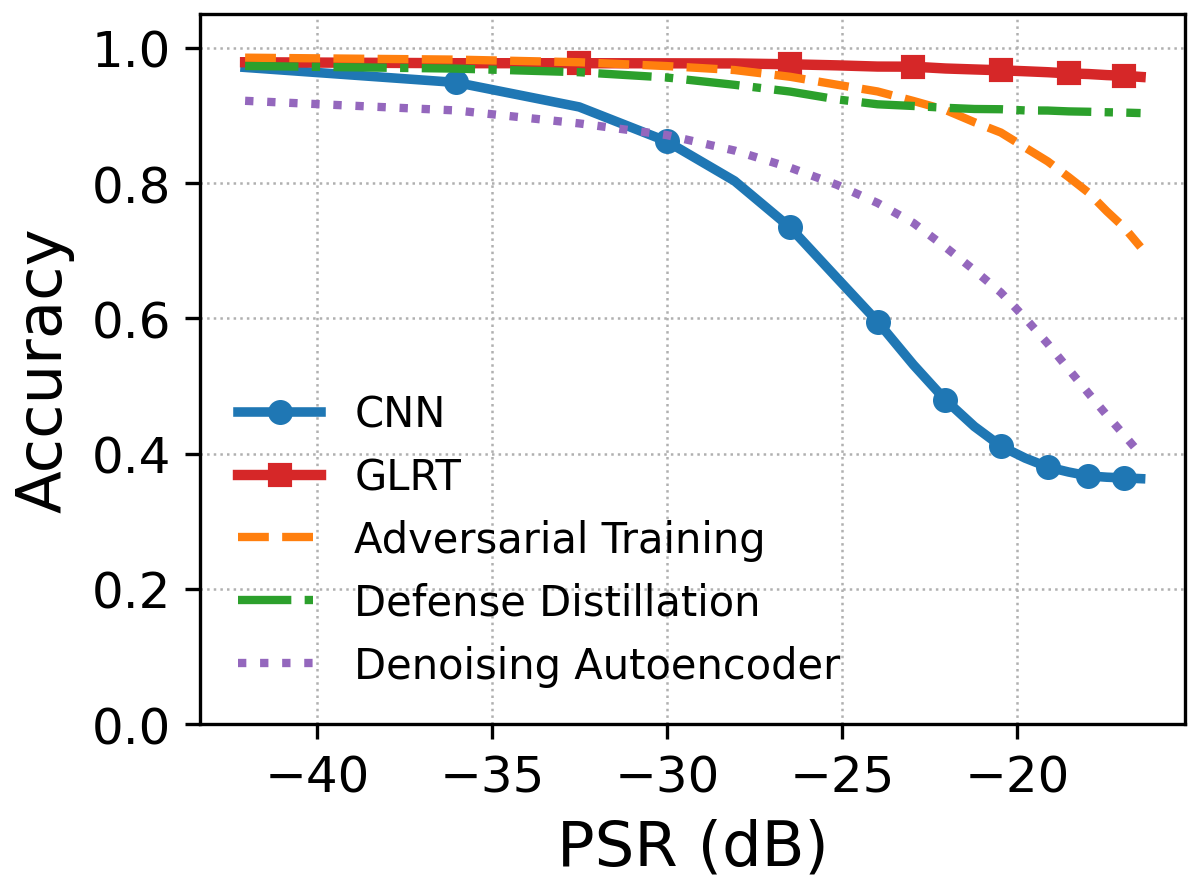}
    }

    \caption{Signal detection performance. The GLRT remains highly robust to adversarial interference. %of our proposed framework in each considered environment. The GLRT demonstrates significantly higher accuracy across multiple PSR levels in comparison to each baseline.
    }
    \label{fig:detection_results}
\end{figure}

\begin{figure}[t] 
    \centering
    % First row
    \subfloat[FGSM $\ell_\infty$]{%
         \includegraphics[width=\halfwidth\columnwidth]{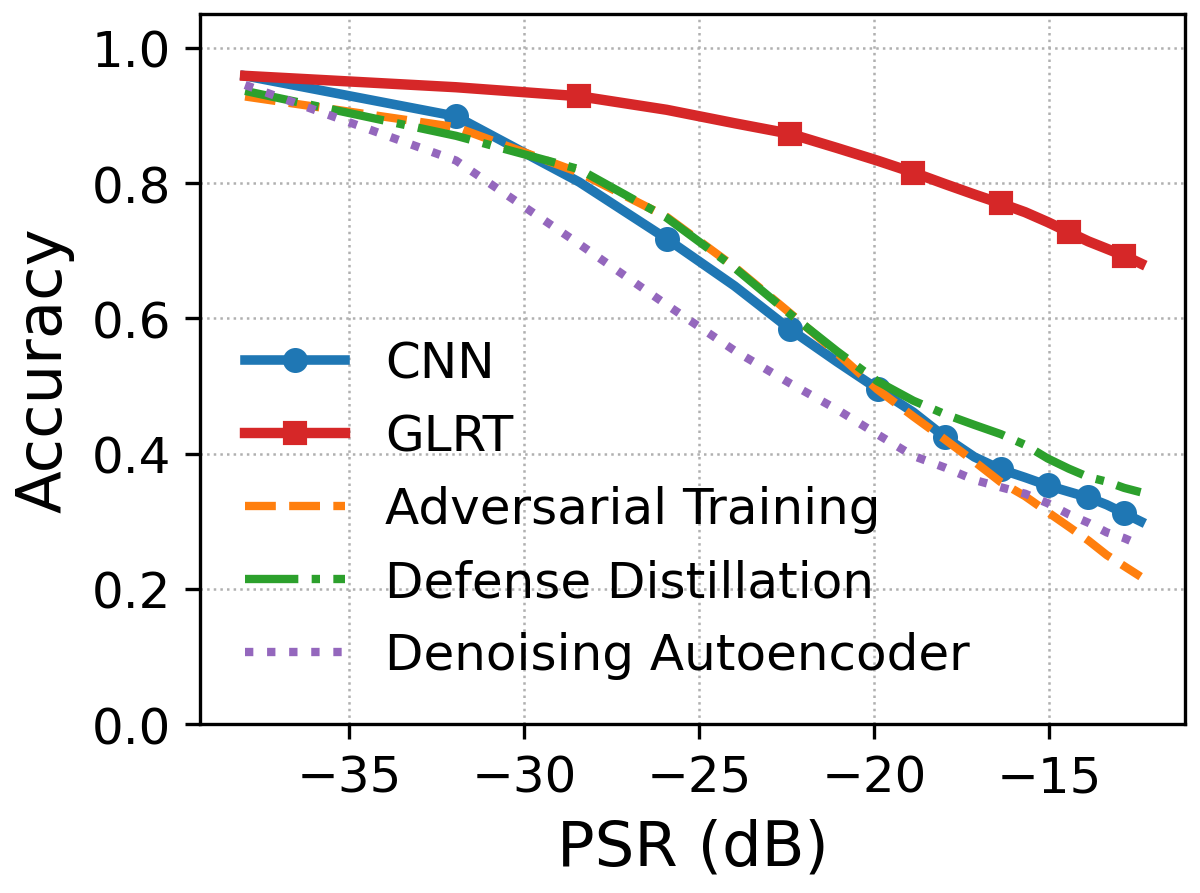}
    }
    \subfloat[FGSM $\ell_2$]{%
         \includegraphics[width=\halfwidth\columnwidth]{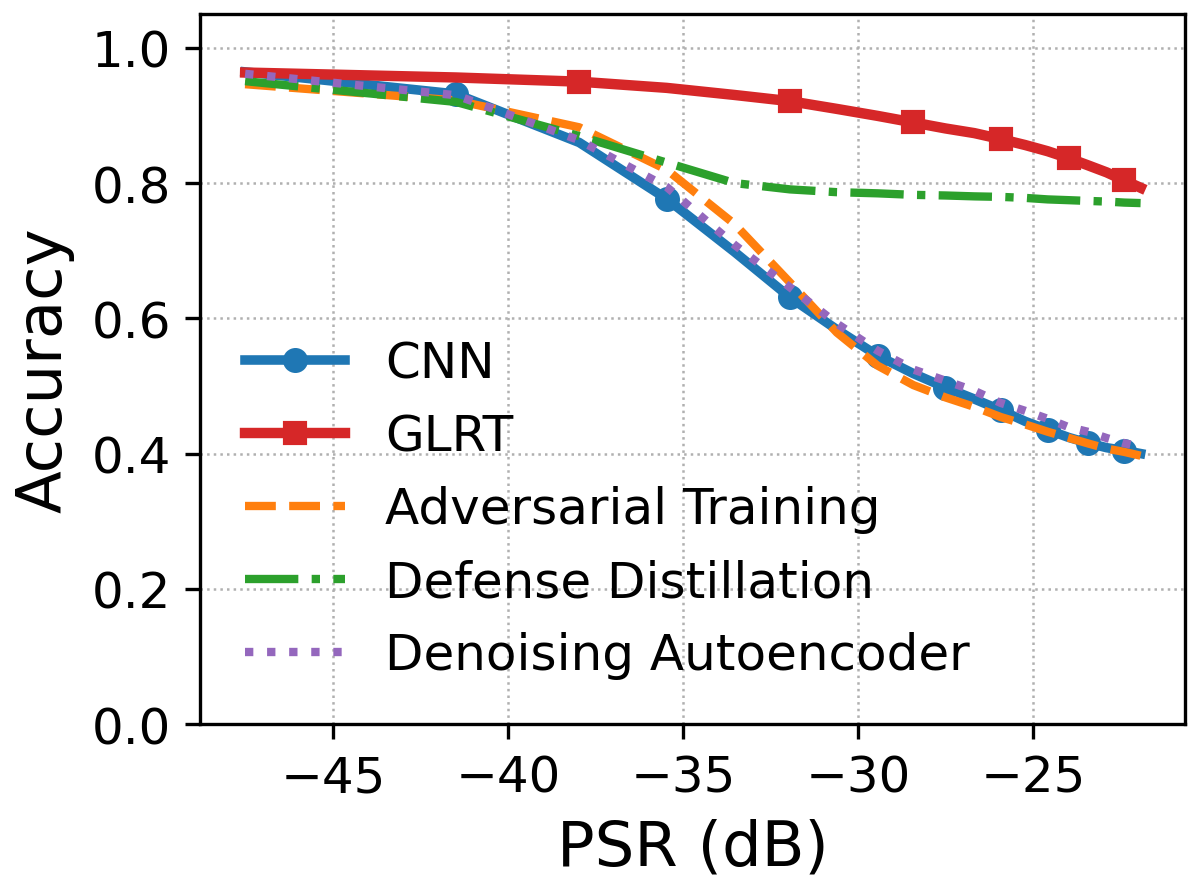}
    }
    
    \subfloat[PGD $\ell_\infty$]{%
        \includegraphics[width=\halfwidth\columnwidth]{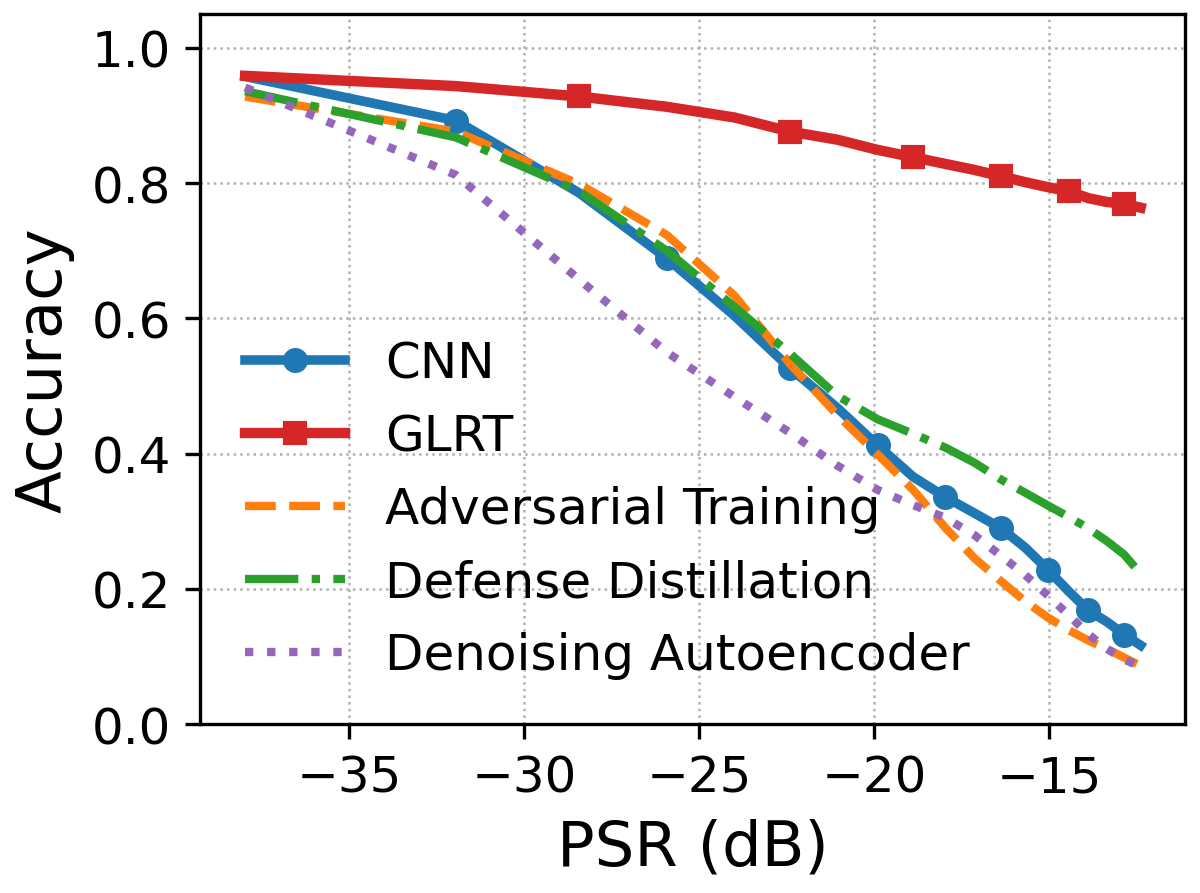}
    }
    \subfloat[PGD $\ell_2$]{%
        \includegraphics[width=\halfwidth\columnwidth]{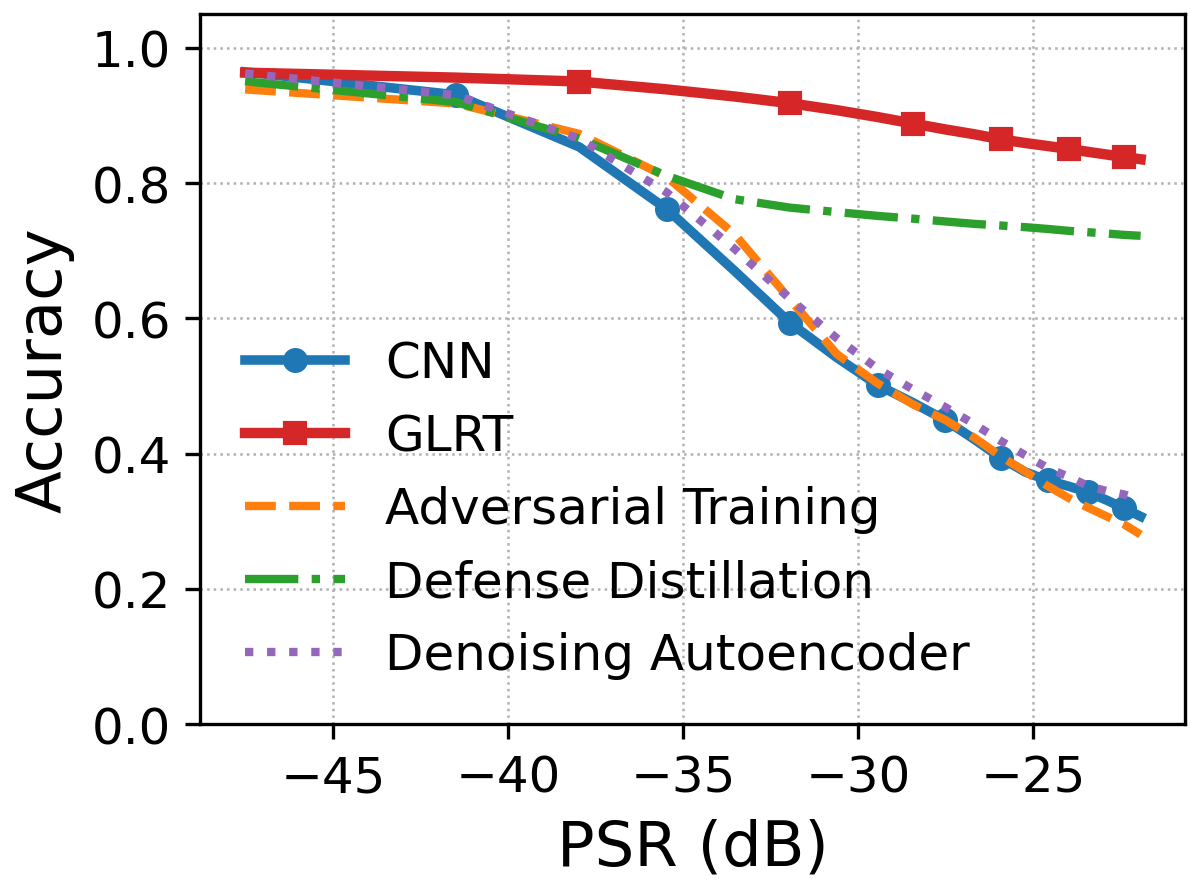}
    }

    \caption{DoA classification performance. Again, the GLRT remains highly robust to adversarial interference.% of our proposed framework in each considered environment. 
    %Similar to Fig. \ref{fig:detection_results}, we see that the GLRT attains higher accuracy in comparison to each baseline.
    }
    \label{fig:doa_results}
\end{figure}

We benchmark our method against three certified state-of-the-art baselines in wireless adversarial machine learning: (i) adversarial training \cite{kim2022advtrain}, which augments the CNN's training set with adversarial examples to reduce gradient sensitivity; (ii) defensive distillation \cite{catak2022defensedistill}, which retrains on softened output probabilities to smooth the loss landscape; and (iii) DAE pre-processing \cite{lee2021dae}, which denoises the input toward the clean-signal manifold before classification. Unlike our framework, all three operate entirely within the DL classifier (or a DL pre-processing stage) and provide only empirical, heuristic robustness, degrading once perturbations exceed the training/calibration regime. In contrast, our GLRT validator is theoretically grounded (Theorem 1) and requires no prior exposure to adversarial examples.

As we see in Figs. \ref{fig:detection_results} and \ref{fig:doa_results}, the CNN accuracy decreases with PSR for both attacks, more severely for DoA (61 classes) than detection (binary); adversarially trained and DAE-defended CNNs partially recover at moderate PSR but collapse at higher PSR, consistent with prior work showing greater effectiveness for $\ell_2$- than $\ell_\infty$-bounded perturbations. The GLRT, in contrast, retains high accuracy throughout, validating the covariance-domain robustness of Theorem \ref{thm:cov-stability}. At very low PSR (near-nominal conditions), the speculative CNN matches GLRT performance, confirming it reliably provides correct speculative estimates on non-adversarial arrays.

\vspace{-0.4cm}
\section{Conclusion}
\label{sec:conclusion}
We theoretically established that the GLRT is highly robust to adversarial interference. Motivated by this result, we developed and empirically validated a speculative framework that integrates a DL classifier with a GLRT estimator to improve robustness against adversarial interference in array processing. Future work will extend our covariance-domain robustness framework to additional DL-based sensing systems such as multi-static ISAC scenarios (e.g., dynamic-TDD and full-duplex settings), where the added self-interference and cross-link channels create further opportunities for adversarial interference in DL-based sensing decisions.

% \textcolor{blue}{Future work will extend this framework to additional DL-based sensing systems such as multi-static ISAC scenarios (e.g., dynamic-TDD and full-duplex settings), where DL introduces adversarial vulnerabilities, where our covariance-domain robustness result (Theorem 1) may similarly apply.} 

%Future work will extend this framework to additional adversarial vulnerabilities (e.g., multiple signal classification) and to multi-static ISAC scenarios (e.g., dynamic-TDD, full-duplex settings), where our covariance-domain robustness result (Theorem 1) may similarly validate DL-based sensing decisions.

\vspace{-0.4cm}
\bibliography{references}

\bibliographystyle{IEEEtran}

\vspace{-0.4cm}
\section*{Appendix: Proof of Theorem \ref{thm:cov-stability}}

% \subsection{Proof of Theorem \ref{thm:cov-stability}}
% \begin{proof}[Proof of Theorem~\ref{thm:cov-stability}]
% \end{proof}
\begin{proof} Given $\mathbf{z}(t)$ for $t = 1, \ldots, T$ as columns of $\mathbf{Z}$ (i.e., $\mathbf{Z} = [\mathbf{z}(1), \mathbf{z}(2), \cdots, \mathbf{z}(T)]$), let $\boldsymbol{\delta}(t)$ for $t = 1, \ldots, T$ be the perturbation of each time sample such that $\boldsymbol{\delta} = [\boldsymbol{\delta}(1), \boldsymbol{\delta}(2), \cdots, \boldsymbol{\delta}(T)]$ and $\|\boldsymbol{\delta}\|_p \le \epsilon$. Define the sample means $\bar{\mathbf{z}}(t) = \frac{1}{T} \sum_{t=1}^T \mathbf{z}(t)$ and $\bar{\boldsymbol{\delta}}(t) = \frac{1}{T} \sum_{t=1}^T \boldsymbol{\delta}(t),$
% \[
% \bar{\mathbf{z}}(t) = \frac{1}{T} \sum_{t=1}^T \mathbf{z}(t), 
% \qquad
% \bar{\boldsymbol{\delta}}(t) = \frac{1}{T} \sum_{t=1}^T \boldsymbol{\delta}(t),
% \]
and the centered data matrices $\mathbf{Z}_c = [\,\mathbf{z}(1) - \bar{\mathbf{z}}(1), \dots, \mathbf{z}(T) - \bar{\mathbf{z}}(T)\,]$ and $\boldsymbol{\delta}_{c} = [\,\boldsymbol{\delta}(1) - \bar{\boldsymbol{\delta}}(1), \dots, \boldsymbol{\delta}(T) - \bar{\boldsymbol{\delta}}(T)\,]$.
The sample covariance matrices are
\begin{equation} \label{z_cov}
    S(\mathbf{Z}) = \frac{1}{T-1} \mathbf{Z}_c \mathbf{Z}_c^{H},
\end{equation}
and
\begin{equation} \label{zdelta_cov}
    S(\mathbf{Z}+\boldsymbol{\delta}) = \frac{1}{T-1} (\mathbf{Z}_c + \boldsymbol{\delta}_c)(\mathbf{Z}_c + \boldsymbol{\delta}_c)^{H}.
\end{equation}
Subtracting (\ref{z_cov}) from (\ref{zdelta_cov}) yields
\begin{equation}
S(\mathbf{Z}+\boldsymbol{\delta}) - S(\mathbf{Z})
= \frac{1}{T-1}
\Big(
\mathbf{Z}_c \boldsymbol{\delta}_c^{H} + \boldsymbol{\delta}_c \mathbf{Z}_c^{H} + \boldsymbol{\delta}_c \boldsymbol{\delta}_c^{H}
\Big).
\label{eq:cov-expansion}
\end{equation}
Now, taking any submultiplicative matrix norm (e.g., the spectral norm $\|\cdot\|_2$), we obtain
\begin{equation}
\|S(\mathbf{Z}+\boldsymbol{\delta}) - S(\mathbf{Z})\|_2
\;\le\; \frac{2}{T-1}\|\mathbf{Z}_c\|_2\,\|\boldsymbol{\delta}_c\|_2
\;+\; \frac{1}{T-1}\|\boldsymbol{\delta}_c\|_2^2.
\label{eq:cov-bound}
\end{equation}
Next, we bound $\|\boldsymbol{\delta}_c\|_2$: $\|\boldsymbol{\delta}\|_p \le \epsilon$ implies $\|\boldsymbol{\delta}_c\|_F^2 \le \sum_t \|\boldsymbol{\delta}(t) - \bar{\boldsymbol{\delta}}(t)\|_2^2 \le c_p T \epsilon^2$, with $c_p = 1$ for $p=2$ and $c_p = M$ for $p=\infty$ (since $\|\boldsymbol{\delta}(t)\|_2 \le \sqrt{M}\,\epsilon$ in that case), so $\|\boldsymbol{\delta}_c\|_2 \le \|\boldsymbol{\delta}_c\|_F \le \sqrt{c_p T}\,\epsilon$. Substituting into (22) gives
\begin{equation}
\|S(\mathbf{Z}+\boldsymbol{\delta}) - S(\mathbf{Z})\|_2 \le \frac{2\sqrt{c_p T}}{T-1}\|\mathbf{Z}_c\|_2\,\epsilon + \frac{c_p T}{T-1}\epsilon^2, \forall p \in \{2,\infty\}. \label{eq:combined_bound}
\end{equation}
The first term is linear in $\epsilon$ and scales with the signal energy $\|\mathbf{Z}_c\|_2$, while the second is quadratic in $\epsilon$; both are negligible when $\epsilon \ll \|\mathbf{Z}_c\|_2/\sqrt{c_p T}$, yielding Theorem 1.

\end{proof}

\end{document}